\documentclass{svjour2}
\usepackage{geometry, amsmath}                
\smartqed  
\usepackage{graphicx}
\usepackage{amssymb}
\usepackage{epstopdf}
\newcommand{\kB}{k_{\text{B}}}
\newcommand{\muex}{\mu^{\text{ex}}}
\newcommand{\xc}{x_{\text{c}}}
\newcommand{\alphac}{\alpha_{\text{c}}}
\newcommand{\vr}{\mathbf{r}}
\newcommand{\vri}{\mathbf{r_i}}

 \journalname{Journal of Statistical Physics}
\begin{document}

\title{Quantifying density fluctuations in volumes of all shapes
and sizes using indirect umbrella sampling}

\author{Amish J. Patel	\and 
	    Patrick Varilly	\and 
	    David Chandler	\and 
	    Shekhar Garde}

\institute{Amish J. Patel \and Shekhar Garde \at
              Howard P. Isermann Department of Chemical \& Biological Engineering, and Center for Biotechnology and Interdisciplinary Studies, Rensselaer Polytechnic Institute, Troy, NY 12180 \\
           \and
           Patrick Varilly \and David Chandler \at
	   Department of Chemistry, University of California, Berkeley, CA 94720 
}

\date{Received: date / Accepted: date
\phantom{\footnotemark}
\footnotetext{To whom correspondence should be addressed. Email: patela10@rpi.edu or chandler@berkeley.edu or  gardes@rpi.edu}
}

\maketitle
\begin{abstract}
Water density fluctuations are an important statistical
mechanical observable that is related to many-body correlations, as well
as hydrophobic hydration and interactions. Local water density
fluctuations at a solid-water surface have also been proposed as a measure of its hydrophobicity. These fluctuations can be quantified by calculating the probability,~$P_v(N)$, of observing $N$~waters in a probe volume of interest~$v$.  When $v$ is large, calculating~$P_v(N)$ using molecular dynamics simulations is challenging, as the probability of observing very few waters is exponentially small, and the standard procedure for overcoming this problem (umbrella sampling in~$N$) leads to undesirable impulsive forces. Patel {\it et al.} [{\it J. Phys. Chem. B}, {\bf 114},
1632 (2010)] have recently developed an indirect umbrella sampling (INDUS) method, that samples a coarse-grained particle number to obtain~$P_v(N)$ in
cuboidal volumes. 
Here, we present and demonstrate an extension of that approach to other basic shapes, like spheres and cylinders, as well as to collections of such volumes. We further describe the implementation of INDUS in the NPT ensemble and calculate $P_v(N)$ distributions over a broad range of pressures. Our method may be of particular interest in characterizing the hydrophobicity of interfaces of proteins, nanotubes and related systems.

\end{abstract}

\section{Introduction}
Quantifying density fluctuations in a condensed phase is interesting
from a statistical physics perspective. For example, the probability~$P_v(N)$ of finding $N$~fluid particles in a probe volume~$v$ contains information about many-body correlations in the fluid. Calculations of $P_v(N)$ in
liquid water have significantly enhanced our understanding of
hydrophobicity.  
In particular, as the hydration of an idealized solvent-excluding hydrophobic solute is equivalent to the creation of a cavity with the same size and shape as that of the solute, the excess free energy, $\muex$, of solute hydration is $-\kB T\,\log P_v(0)$~\cite{widom_jcp63}. 
In 1996, Hummer {\it et al.} showed that in bulk water, $P_v(N)$ distributions are gaussian for small spherical volumes containing fewer than ten water molecules on average \cite{information_theory}. This simplicity formed the basis for an information theoretic model that could predict the
thermodynamics of hydrophobic hydration and the association of small
solutes over a range of conditions, using only the readily available information
on the average density and the water radial distribution function
\cite{information_theory,garde96,hummer98pnas}. Gaussian statistics of density fluctuations \cite{gaussian_ft} also underlies the Pratt-Chandler theory~\cite{pratt_chandler}, which employs the same information to estimate pair correlation functions for small hydrated hydrophobic species.

While small solutes can be accommodated in cavities that are formed
spontaneously by thermal fluctuations in bulk water, solvating large
solutes requires forming a liquid-vapor-like interface
\cite{stillinger,LCW,DC_nature05}.  As a result, the nature of density
fluctuations in large volumes is more complex. The
Lum-Chandler-Weeks (LCW) theory captures the lengthscale dependence of
hydration quantitatively by combining the physics of gaussian
density fluctuations and that of interface formation~\cite{LCW}.
Specifically, it predicts that while $P_v(N)$ for large volumes is
gaussian around the mean, the low-$N$ wings of the distribution are enhanced substantially~\cite{lum_thesis,LLCW}.  
Quantifying these rare water fluctuations in large volumes is essentially impossible in equilibrium molecular simulations, and requires non-Boltzmann or umbrella sampling methods~\cite{DCbook}. 
Straightforward umbrella sampling of $N$, is further complicated by the fact that $N$ is a discontinuous function of particle coordinates, resulting in impulsive forces, which are difficult to treat in typical molecular dynamics (MD) simulations. To circumvent this difficulty, Patel {\it et al.}  recently introduced an indirect umbrella sampling (INDUS) method in which $N$ is sampled indirectly, by biasing a coarse-grained variable,~$\tilde N$, which is strongly
correlated with $N$ but varies continuously with particle coordinates~\cite{ajp_jpcb2010}. The original implementation of INDUS is suitable only for cuboidal volumes, and showed that for large volumes in bulk water, 
$P_v(N)$ indeed deviates significantly from gaussian behavior at low~$N$, reflecting the underlying physics of interface formation~\cite{ajp_jpcb2010}.  

Application of INDUS to sample density fluctuations in large
volumes in interfacial environments showed that fluctuations near
hydrophilic surfaces are similar to those in bulk, but near
hydrophobic interfaces, the probability of density depletion is
significantly enhanced~\cite{ajp_jpcb2010}. The ability to calculate $P_v(N)$, and especially $\muex=-\kB T\,\log P_v(0)$, in large volumes near interfaces also allowed us to calculate the binding free energies of hydrophobic cuboids to surfaces with a range of chemistries~\cite{AJP_Lscale}, and these binding free energies were shown to correlate with the macroscopic wetting properties of the surfaces. Thus, $P_v(N)$ is a potential molecular measure of hydrophobicity, which may enable the characterization of surfaces of proteins and biomolecules that exhibit nanoscale roughness and chemical heterogeneity~\cite{AJP_Lscale,garde09pnas,acharya:faraday,garde09prl}.

Here, we extend INDUS such that it can be used to umbrella
sample probe volumes of other regular shapes, {\it e.g.}, with cylindrical
and spherical symmetry, as well as intersections and unions of collections of such regular volumes and their complements. While the ideas underlying the extension are simple, they considerably widen the scope of the method. For example, they allow umbrella sampling of arbitrarily shaped volumes, enabling faithful characterization of flucuations in the hydration shells of ions, nanoparticles, nanotubes, and the rugged surfaces of proteins.

We also extend the method to work in the NPT ensemble. 
Previous applications of INDUS were performed in the NVT ensemble with a
buffering vapor-liquid interface.  
While the two schemes yield indistinguishable results at low pressures, the present extension allows access to a much broader range of pressures.
We begin by describing the INDUS method of Ref.~\cite{ajp_jpcb2010},
which is suitable for cuboidal probe volumes, and introduce the
pertinent equations, which lays down the framework for extending the
method to other regular volumes. 
We then generalize these equations to volumes of more general shapes and to collections of such volumes, and describe how INDUS affects the calculation of system pressure. 
Finally, we demonstrate these generalizations by calculating $P_v(N)$
in various noncuboidal shapes and at high pressures.

\section{The INDUS Method}
The number of particles, $N$, in a specific probe volume, $v$, changes
discontinuously as the center of any particle crosses the surface of
$v$. Hence, if the biasing potential, $U$, were chosen to be a
function of $N$, it would result in impulsive forces. Instead, we
choose $U$ to be a function of a closely related coarse-grained
particle number, ${\tilde{N}}$, that is a continuous function of the positions, $\{\mathbf{r_i}\}$, of all $M$ particles in the system as,
\begin{align*}
\tilde{N}&=\sum_{i=1}^{M}\tilde{h}(\mathbf{r_i}), \text{\ where}
\tag{1a}\\
\quad\tilde{h}(\mathbf{r_i})&\equiv\int_v   \Phi(\mathbf{r}-\mathbf{r_i})\,\text{d}\mathbf{r}. 
\tag{1b}
\end{align*}
The integral in Eq. 1b is over the probe volume $v$, and the integrand is a coarse-graining function, $\Phi(\mathbf{r})$, which we choose to be
\begin{align*}
\Phi(\mathbf{r}) &= \phi(x)\phi(y)\phi(z),\text{\ where}
\tag{2a} \\
\quad\phi(\alpha)&= k^{-1} [ e^{-\alpha^2 / 2\sigma^2} - e^{-\alphac^2 / 2\sigma^2} ] \Theta( \alphac - |\alpha| ).
\tag{2b}
\end{align*}
The function $\phi(\alpha)$, shown in Figure~1, is a gaussian that is truncated at $|\alpha|=\alphac$, shifted down, and then scaled, so as to make it continuous and normalized. The normalization constant, $k$, is equal to
$\sqrt{2\pi\sigma^{2}}\,\text{erf}(\alphac/\sqrt{2\sigma^2})-2\alphac\exp(-\alphac^2 / 2\sigma^2)$~and $\Theta(\alpha)$ is the Heaviside step function.
As the width of the gaussian, $\sigma$, approaches~$0$, the function~$\phi(\alpha)$ approaches the Dirac delta function~$\delta(\alpha)$ and ${\tilde N}$
approaches~$N$.  The correlation between $\tilde N$ and $N$ is thus strongest when $\sigma$ is smallest, but if $\sigma$ is chosen to be too small, the resulting biasing forces may be too large to handle correctly in typical MD simulations.

\begin{figure}[htbp] 
   \centering
   \includegraphics[width=2.in]{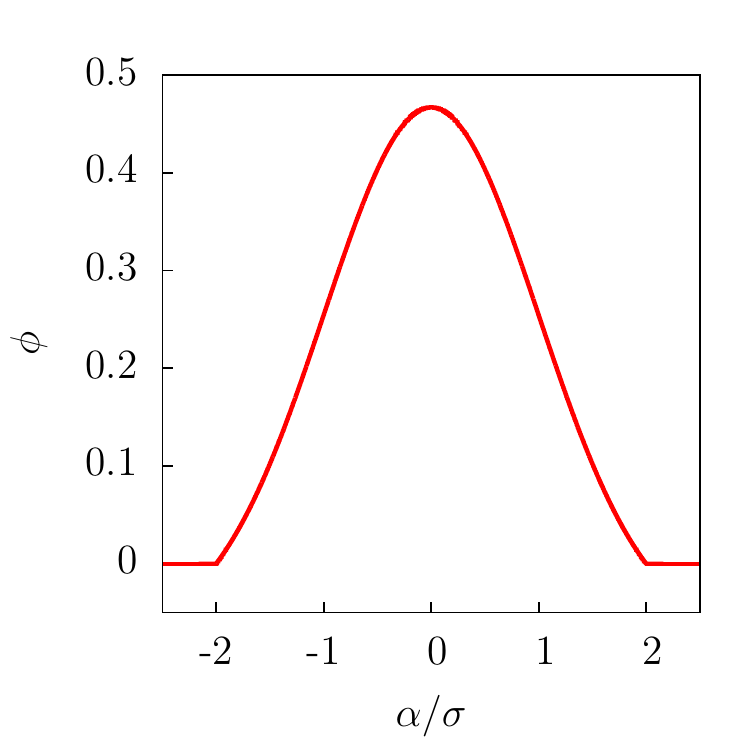} 
   \caption{Coarse-graining function, $\phi(\alpha)$, as defined in Eq. 2b, for $\alphac=2\sigma$.}
   \label{fig:phix}
\end{figure}

For a cuboidal volume~$v$, the integral in Eq. 1b can be
performed independently in the $x$, $y$~and~$z$ directions.  The result is
\begin{align*}
\tilde{h}(\mathbf{r_i}) &=
\tilde{h}_{x}(x_i)\tilde{h}_{y}(y_i)\tilde{h}_{z}(z_i),\text{\ where}
\tag{3a} \\
\quad\tilde{h}_{x}(x_i)&=
\int_{x_{\mathrm{min}}}^{x_{\mathrm{max}}} \phi(x-x_i)\,\text{d}x,
\tag{3b} 
\end{align*}
and $x_{\mathrm{min}}$~and~$x_{\mathrm{max}}$ are the coordinates of the faces of $v$ perpendicular to the $x$-axis.  The functions $\tilde h_y(y_i)$~and~$\tilde h_z(z_i)$ are defined analogously.

\begin{figure}[htbp] 
   \centering
   \hspace{-0.24in}\includegraphics[width=2.22in]{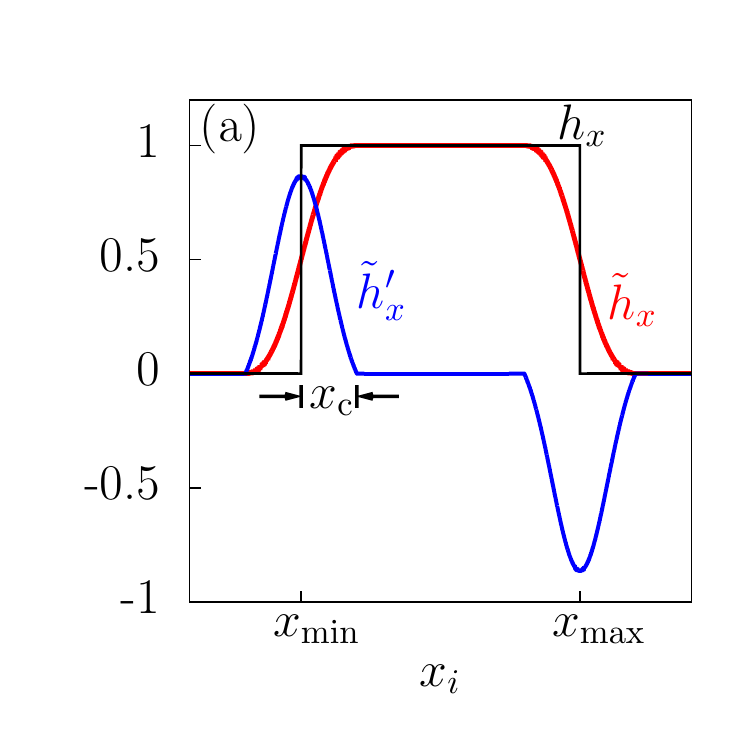}\hspace{-0.24in}\includegraphics[width=2.22in]{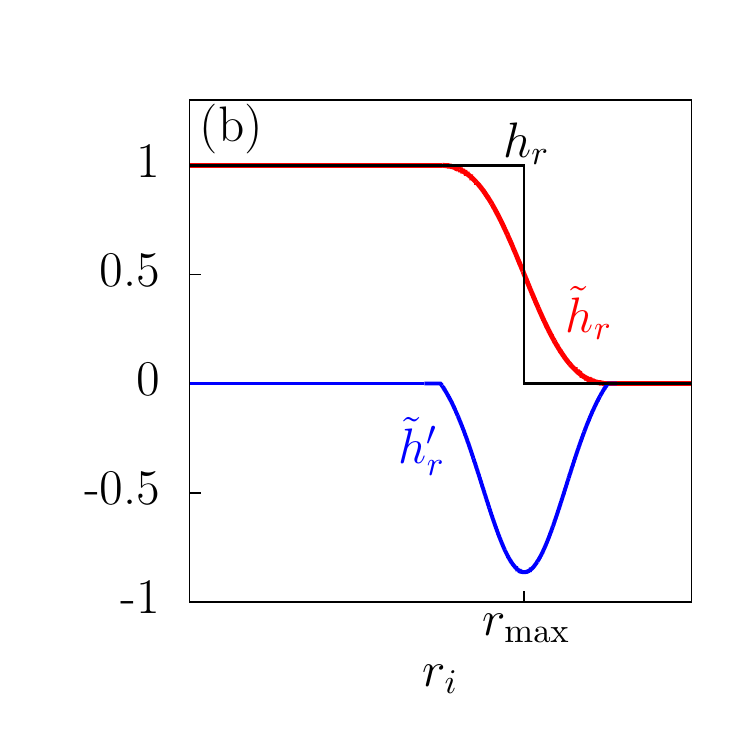}\hspace{-0.24in}\includegraphics[width=2.22in]{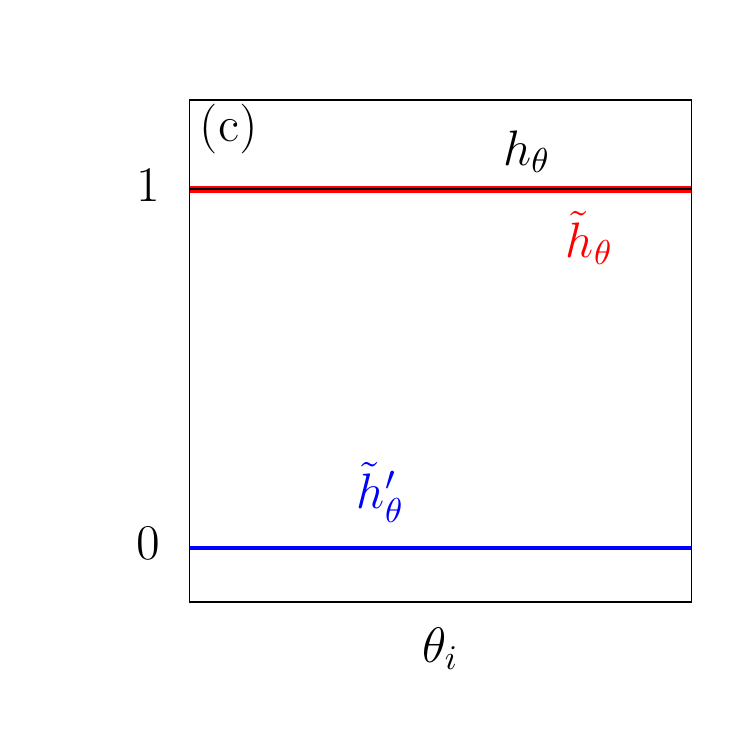}  
   \caption{The functions $h_{\alpha}(\alpha_i)$, $\tilde{h}_{\alpha}(\alpha_i)$ and its derivative, $\tilde{h}'_{\alpha}(\alpha_i)$, for coordinates that have (a) two ($\alpha\to x$), (b) one ($\alpha\to r$) or (c) zero ($\alpha\to\theta$) boundaries.}
   \end{figure}

Fig. 2a shows the function~$h_x(x_i)$ (equal to $1$ for $x_{\rm min}\le x_i \le x_{\rm max}$, and $0$ otherwise), which can be thought of as the $x$~contribution to $h(\vri)$; that is, $h(\vri) = h_x(x_i) h_y(y_i) h_z(z_i)$ and $N=\sum_i h(\vri)$.
Fig. 2a also shows the function~$\tilde{h}_x(x_i)$, which
varies continuously across the boundary of~$v$, unlike $h_x(x_i)$. The
coarse-graining function $\tilde{h}_x(x_i)$ differs from $h_x(x_i)$ only
in the thin boundary region of thickness~$2 \xc$. Thus, by ensuring that $\tilde{N}$~and~$N$ are strongly correlated, we are able to influence $N$ indirectly by biasing~$\tilde{N}$.

For a cuboidal probe volume, the $x$-component of the force on particle~$i$ due to the biasing potential, $U(\tilde{N})$, is given by
\begin{align*}
f_{x,i}\equiv-\frac{\partial U}{\partial x_{i}}
=-\frac{\partial U}{\partial\tilde{N}}\frac{\partial\tilde{h}(\vri)}{\partial x_i}=-\frac{\partial U}{\partial \tilde{N}} \tilde h'_x(x_{i}) \tilde h_y(y_{i}) \tilde h_z(z_{i}), 
\tag{4}
\end{align*}
where the derivative of $\tilde{h}_{x}(x_i)$, obtained by
differentiating Eq.~3b and shown in Fig.~2a, is
\begin{align*}
\tilde{h}_{x}'(x_i)
=-[\phi(x_{\mathrm{max}}-x_i)-\phi(x_{\mathrm{min}}-x_i)].
\tag{5}
\end{align*}
It follows that the biasing forces act only on particles near the boundary of~$v$, are finite, and are continuous functions of particle positions.

To obtain $P_v(N)$ using INDUS, we perform $n_w$ simulations with different biasing potentials, $U_j(\tilde{N})$ ($j=1,\ldots,n_w$), chosen such that the range of interest of $N$ is well sampled. During each simulation, we collect $n_j$ samples of $N$ and $\tilde{N}$, denoted by $N_{j,l}$ and
$\tilde{N}_{j,l}$ ($l=1,\ldots,n_j$), in essence, sampling the biased joint distribution function, $P_v(N,\tilde{N})$.
We then unbias and stitch together the $n_w$ biased joint distribution functions by using the weighted histogram analysis method (WHAM)~\cite{wham2,roux_wham}.
Finally, we integrate out the unbiased joint distribution function to obtain $P_v(N)$, which is given by
\begin{align*}
P_v(N) = C
&\sum_{j=1}^{n_w}\sum_{l=1}^{n_j}\frac{\delta_{N,N_{j,l}}}{\sum_{i=1}^{n_w}
n_i e^{-\beta[U_i(\tilde{N}_{j,l})-c_i]}} ,
\tag{6}
\end{align*}
where $\delta_{n,m}$ is the Kronecker delta function, and $C$ and
$\{c_j\}$ are normalization constants.  These are chosen self-consistently via the standard WHAM equations,
\begin{align*}
C^{-1} =
&\sum_{j=1}^{n_w}\sum_{l=1}^{n_j}\frac{1}{\sum_{i=1}^{n_w}
n_i e^{-\beta[U_i(\tilde{N}_{j,l})-c_i]}}, {~\rm and}  
\tag{7a}\\
e^{-\beta c_k} =C
&\sum_{j=1}^{n_w}\sum_{l=1}^{n_j}\frac{e^{-\beta U_k(\tilde N_{j,l})}}{\sum_{i=1}^{n_w}
n_i e^{-\beta[U_i(\tilde{N}_{j,l})-c_i]}}. 
\tag{7b}
\end{align*}

\section{Extension of INDUS to noncuboidal volumes}
While several coarse-graining schemes are possible for defining
$\tilde N$, a practically useful definition must satisfy the following three conditions: 
(i) $\tilde N$ must be a continuous function of particle positions, (ii) $\tilde N$~and~$N$ must be strongly correlated, and (iii) the calculation of~$\tilde N$
and its derivatives should be straightforward.
The choice of the form of Eq. 2a for cuboid volumes allows $\tilde h(\vri)$ to
be expressed as a product of independent contributions from $x$, $y$,
and $z$ coordinates (as in Eq. 3a). 
While this formulation is particularly convenient for cuboidal volumes, the integral (Eq.~1b) that defines $\tilde{h}(\vri)$ would not be independent in the three coordinates for other regular volumes, such as spheres or cylindrical
shells. Thus, calculating $\tilde h(\vri)$ and its gradient efficiently at every MD step would not be straightforward. To circumvent this complication, we bypass defining $\tilde{h}(\vri)$ via a coarse-graining function $\Phi$ as in Eq.~1b, and instead, define it directly as a product of independent contributions from the three co-ordinates (as in Eq.~3a) in the relevant co-ordinate system ({\it e.g.}, cylindrical, spherical, etc.) as,
\begin{align*}
\tilde{h}(\vri) &= \prod_{\alpha}\tilde{h}_{\alpha}(\alpha_i).
\tag{8}
\end{align*}
Here $\alpha$ represents the coordinates component index ($x$, $y$~or~$z$ in Cartesian coordinates; $r$, $\theta$~or~$z$ for cylindrical ones, etc.) and
$\tilde h_\alpha(\alpha_i)$ may be defined in a manner analogous to
$\tilde h_x(x_i)$ (Eq. 3b and Fig. 2a).

However, unlike cuboidal volumes, where each coordinate component has two boundaries ({\it e.g.}, $x_{\mathrm{min}}$ and $x_{\mathrm{max}}$), the components in spherical or cylindrical systems may have either one boundary ({\it e.g.}, the $r$~coordinate for a spherical $v$), or no boundaries ({\it e.g.}, the $\theta$~coordinate for a cylindrical $v$). These cases are illustrated in Fig.~2 and the expressions for $\tilde h_\alpha(\alpha_i)$ and $\tilde h'_\alpha(\alpha_i)$ in each case are as follows:
\begin{itemize}
\item \textbf{Two boundaries:} $\alpha_{\text{min}} \le \alpha \le \alpha_{\text{max}}$.
\begin{align*}
\tilde{h}_{\alpha}(\alpha_i)&=
\bigg[k_1\mathrm{erf}\bigg(\frac{\alpha_{\mathrm{max}}-\alpha_i}{\sqrt{2}\sigma}\bigg) - k_2(\alpha_{\mathrm{max}}-\alpha_i)-\frac{1}{2}\bigg]\Theta(\alphac-|\alpha_{\mathrm{max}}-\alpha_i|) \\
&+
\bigg[k_1\mathrm{erf}\bigg(\frac{\alpha_i-\alpha_{\mathrm{min}}}{\sqrt{2}\sigma}\bigg) - k_2(\alpha_i-\alpha_{\mathrm{min}})-\frac{1}{2}\bigg]\Theta(\alphac-|\alpha_i-\alpha_{\mathrm{min}}|) \\
&+
\Theta\bigg(\alphac+\frac{1}{2}(\alpha_{\mathrm{max}}-\alpha_{\mathrm{min}})-\bigg|\alpha_i - \frac{1}{2}(\alpha_{\mathrm{min}}+\alpha_{\mathrm{max}})\bigg|\bigg),~\text{and}
\tag{9a}\\
\tilde{h}_{\alpha}'(\alpha_i)
&= -[\phi(\alpha_{\mathrm{max}}-\alpha_i)-
\phi(\alpha_{\mathrm{min}} - \alpha_i)],
\tag{9b}
\end{align*}
where $k_1=k^{-1}\sqrt{\pi\sigma^2/2}$ and $k_2=k^{-1}\exp(-\alphac^2/2\sigma^2)$.
\vspace*{\baselineskip}
\item \textbf{One boundary:} $\alpha \le \alpha_{\text{max}}$.
\begin{align*}
\tilde{h}_{\alpha}(\alpha_i)&=
\bigg[k_1\mathrm{erf}\bigg(\frac{\alpha_{\mathrm{max}}-\alpha_i}{\sqrt{2}\sigma}\bigg) - k_2(\alpha_{\mathrm{max}}-\alpha_i) - \frac{1}{2}\bigg]\Theta(\alphac-|\alpha_{\mathrm{max}}-\alpha_i|) \\
&+
\Theta(\alphac+\alpha_{\mathrm{max}}-\alpha_i),~\text{and}
\tag{10a}\\
\tilde{h}_{\alpha}'(\alpha_i)
&= -\phi(\alpha_{\mathrm{max}}-\alpha_i).
\tag{10b}
\end{align*}
\item \textbf{No boundaries:}
\begin{align*}
\tilde{h}_{\alpha}(\alpha_i)&= 1,\quad\text{and}
\tag{11a}\\
\tilde{h}_{\alpha}'(\alpha_i) &= 0.
\tag{11b}
\end{align*}
\end{itemize}
The forces are then given by
\begin{align*}
f_{x,i} &= -\frac{\partial U}{\partial \tilde{N}}\frac{\partial\tilde{h}(\vri)}{\partial x_i}
\tag{12a}, ~\text{with} \\
\frac{\partial\tilde{h}(\vri)}{\partial x_i} &=
\sum_{\alpha}\Biggl[
\tilde{h}'_{\alpha}(\alpha_i)\frac{\partial \alpha_i}{\partial x_i}\,\prod_{\gamma\ne\alpha}\tilde{h}_{\gamma}(\gamma_i)
\Biggr],
\tag{12b}
\end{align*}
where $\partial \alpha_i / \partial x_i$ is an element of the
Jacobian for the coordinate transformation.  \\

\section{Generalization to collections of probe volumes}
The above approach can be generalized to calculate $P_v(N)$ in a
probe volume~$v$ that is constructed from unions ($v_A\cup v_B$) and intersections ($v_A\cap v_B$) of regular subvolumes ($v_A,v_B$) and their complements ($v_{A'},v_{B'}$). The subvolumes need not be of the same size or shape. When $v$ is constructed from subvolumes using the complement, intersection and union operations, the corresponding definition of $\tilde{h}(\vri)$ is constructed by noting that,
\begin{align*}
\tilde{h}^{(A')} &= 1-\tilde{h}^{(A)},
\tag{13a} \\
\tilde{h}^{(A\cap B)} &= \tilde{h}^{(A)}\tilde{h}^{(B)},~\rm{and}
\tag{13b} \\
\tilde{h}^{(A\cup B)} &= 1-\tilde{h}^{(A')}\tilde{h}^{(B')}.
\tag{13c}
\end{align*}
Here, the superscript~$(A)$ indicates that the function is evaluated with respect to the boundaries of sub-volume~$v_A$. 
For the special case of a probe volume~$v$ that is a union of $G$ 
non-overlapping sub-volumes~$\{v_k\}$ ($k=1,\ldots, G$), the above prescription yields,
\begin{align*}
\tilde{h}(\vr_i) &= \sum_{k=1}^{G}\tilde{h}^{(k)}(\vr_i),~\rm{where}
\tag{14a} \\
\tilde{h}^{(k)}(\vri) &=\prod_{\alpha}\tilde{h}_{\alpha}^{(k)}(\alpha_i).
\tag{14b}
\end{align*}

Once again, the force on particle $i$ resulting from a biasing
potential, $U$, is finite and continuous everywhere, and is given by
\begin{align*}
&f_{x,i} = 
-\frac{\partial U}{\partial\tilde{N}}\frac{\partial \tilde{N}}{\partial x_{i}},~\rm{where}
\tag{15a} \\
&\frac{\partial \tilde{N}}{\partial x_{i}}
=\sum_{k=1}^{G}\frac{\partial \tilde{h}^{(k)}(\vri)}{\partial x_i}.
\tag{15b}
\end{align*}
The recipe given in Eqs. 9-12, when applied to $v_k$ can be used to evaluate
$\tilde h^{(k)}_{\alpha}$ and $\partial\tilde{h}^{(k)}/\partial x_i$ in Eqs. 14b and 15b.

\section{INDUS in the NPT ensemble}

When calculating $P_{v}(N)$ using simulations in the NVT
ensemble, as was done in Ref.~\cite{ajp_jpcb2010}, it is important to have a vapor bubble or a vapor-liquid interface in the simulation box. This vapor bubble can be nucleated, {\it e.g.}, by applying a particle excluding field far from $v$, and can grow or shrink to accommodate water molecules pushed into or out of~$v$.  The resulting effective pressure of the system is close to the saturation vapor pressure of the fluid. Alternatively, we can perform simulations in the NPT ensemble without such a bubble, as long as the forces resulting from the umbrella potential are included in the calculation of the system pressure, ${\mathcal P}$. If~$v$ is fixed in space and does not move, grow or shrink as the simulation box dimensions fluctuate, then the contribution of the umbrella potential to ${\mathcal P}$ is
\begin{align*}
\mathcal{P}^{\rm umb} \equiv -\frac{\partial U}{\partial V}=\frac{1}{3V}\sum_{i=1}^{M} \mathbf{r}_i\bullet\mathbf{f}_i^{\rm umb},
\tag{16}
\end{align*}
where $\mathbf{f}_i^{\rm umb}$ is the umbrella force on particle $i$, calculated as described in the preceding sections, and $V$ is the system volume.


\section{Results}

 \begin{figure}[htbp] 
   \centering
   \includegraphics[width=3.in]{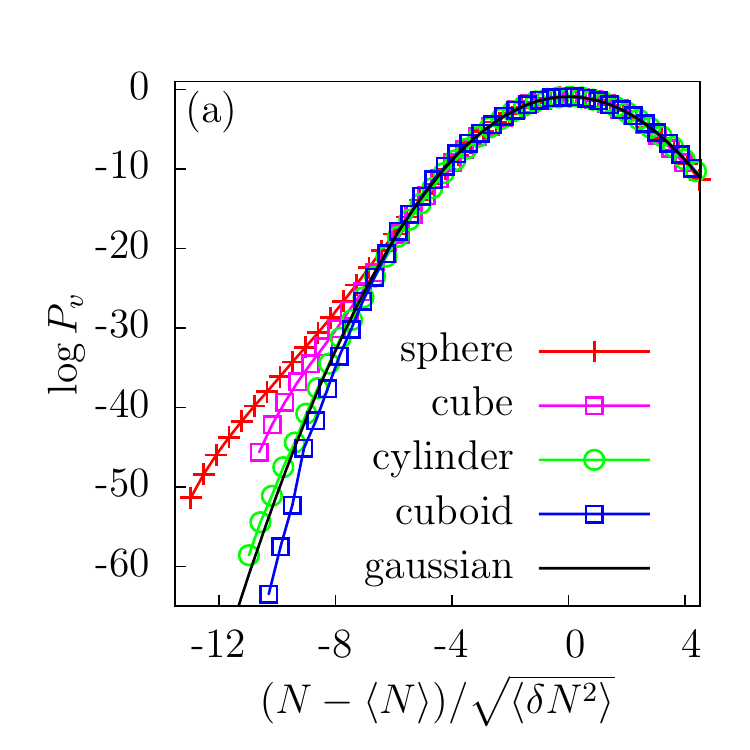}\includegraphics[width=3.in]{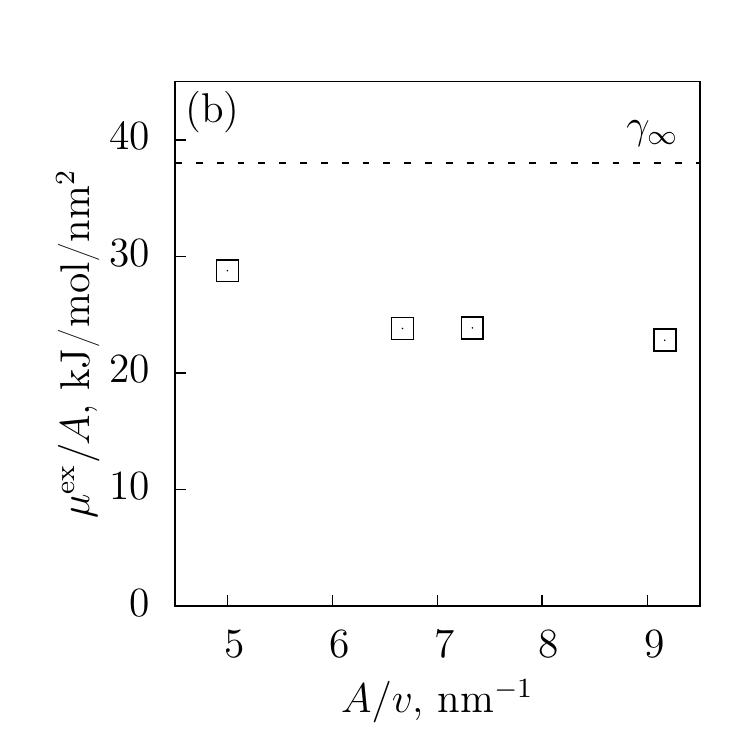} 
   \caption{(a) $\log P_v$ as a function of $(N-\langle
   N\rangle)/\sqrt{\langle \delta N^2\rangle}$ for volumes of four
   different shapes: a sphere of radius~$0.6\,$nm, a cube of side~$0.9\,$nm, a
   cylinder of radius~$0.3\,$nm and length~$3\,$nm, and a thin cuboid of dimensions $0.3\,\text{nm}\times1.6\,\text{nm}\times1.6\,\text{nm}$. (b) The ratio of $\mu^{\rm ex}$ to surface area~$A$, as a function of $A / v$ for the four different shapes. The dashed line represents the surface tension,~$\gamma_{\infty}$, of a vapor-liquid interface of SPC/E water~\cite{VegaMiguel2007}.}
   \end{figure}

We illustrate the utility of the INDUS method by calculating $P_v(N)$
distributions for volumes of different shapes in bulk water. Biased MD simulations of bulk water were performed using the packages LAMMPS and GROMACS~\cite{lammpsref,gmx4ref}, modified in-house to implement INDUS. For the parameters of the coarse-graining function $\phi(\alpha)$ in Eq.~2b, we used $\sigma=0.1\,$\AA\  and $\alphac=0.2$\,\AA\ (NVT ensemble) or $\alphac=0.3$\,\AA\ (NPT ensemble).
Each simulation box used contained several thousand water molecules, 
modeled with the extended simple point charge water model (SPC/E)~\cite{spce}, and was periodic in all directions.

We selected volumes of four different shapes (a sphere, a cube, a cylinder, and a cuboid; see Figure 3), each with an average number of water molecules, $\langle N \rangle$, between $25$~and~$30$. For
these large volumes, INDUS allows us to measure probabilities for rare
water fluctuations that are rather small ($P_v(0)\approx 10^{-30}$), whereas calculations using straightforward equilibrium simulations~\cite{information_theory} provide accurate estimates only for much smaller volumes ($\langle N \rangle \approx 8$ with corresponding
$P_v(0)\approx 10^{-8}$).

Although the volumes of the shapes that we have selected are similar to each
other, they are not identical. Therefore, to compare them, in Figure
3a, we plot $P_v$ as a function of $(N - \langle N \rangle
)/\sqrt{\langle \delta N^2 \rangle}$, where $\langle \delta N^2
\rangle$ is the variance of $N$. Near the mean, fluctuations are
gaussian for all shapes, as expected. However, there are deviations from such gaussian behavior in the tails of $P_v(N)$. Specifically, the smaller a shape's surface-area to volume ratio, the fatter the low-$N$ tail.

In the large lengthscale limit, interface formation governs the free
energy of cavity formation. LCW theory~\cite{LCW} predicted, and subsequent
simulation studies verified~\cite{HGC,garde05,ashbaugh_SPT}, that the
gradual crossover from small to large lengthscale physics occurs around $1\,$nm, which is roughly the lengthscale of volumes selected
here. Thus, we expect that shapes with smaller surface areas will have
lower free energies of cavity formation and correspondingly fatter
low $N$ tails, as observed in Figure~3a. Figure~3b further confirms
that the free energy is governed by the physics of interface formation: the ratio of $\mu^{\rm ex}$ to the surface area of the probe volume, $A$, which can be interpreted as an apparent surface tension for these nanoscopic objects, is approximately constant, independent of the shape of~$v$.  This apparent surface tension is lower than the surface tension of a vapor-liquid interface,
consistent with results of Patel {\it et al.} \cite{ajp_jpcb2010}.

\begin{figure}[htbp] 
   \centering
   \includegraphics[width=3.0in]{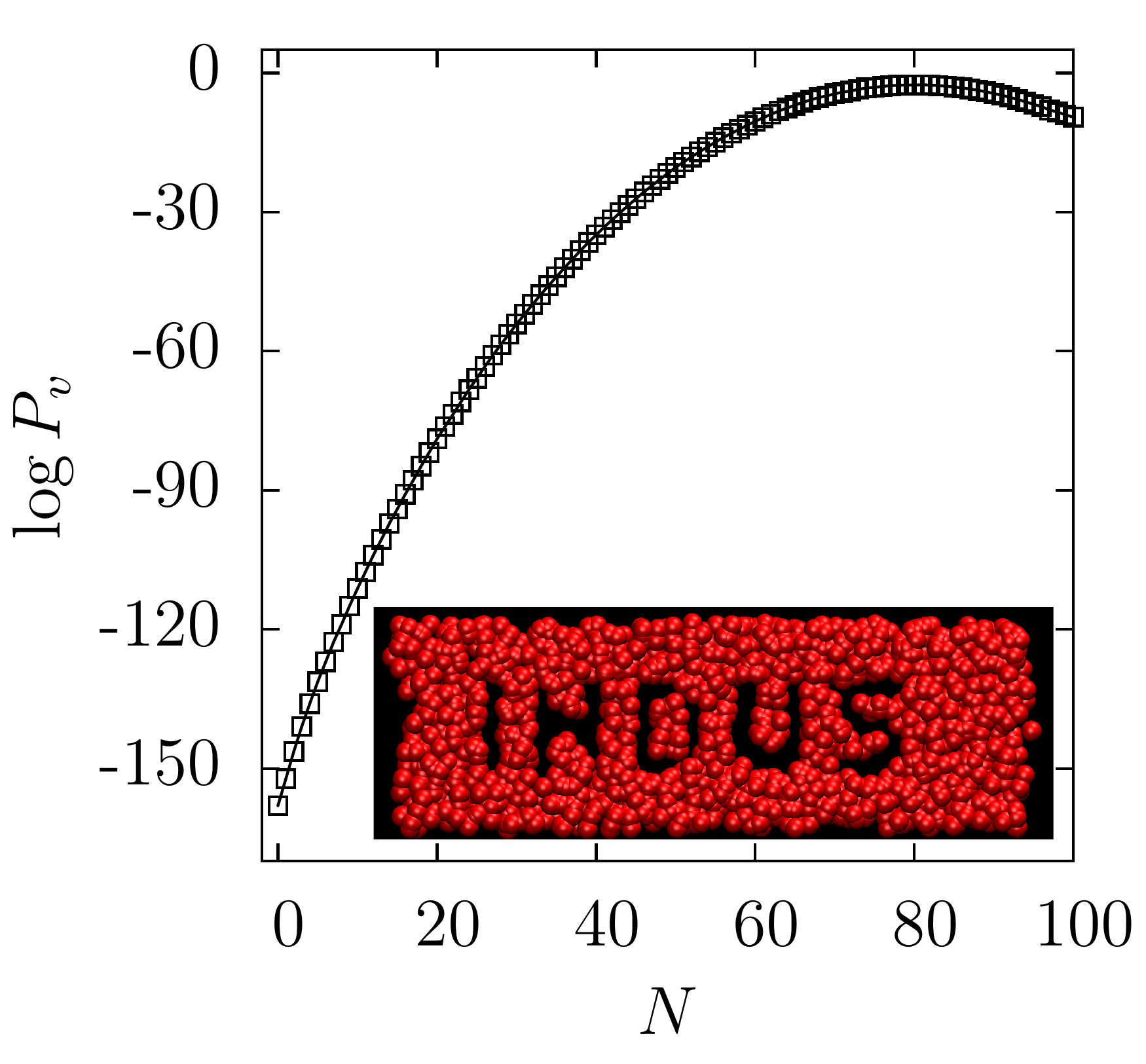}
   \caption{$P_v(N)$ obtained by umbrella sampling a probe volume
   that spells, `I N D U S'. The volume is composed of 156 cubic subvolumes of side~$0.25\,$nm. The inset shows a superposition of five independent
   configurations, taken from an MD simulation with a strong biasing
   potential that empties the probe volume. The red spheres represent
   water oxygens. The letter `I' in the inset is $0.5\,$nm wide and $2.0\,$nm tall.
}
\vspace{-0.2in}
\end{figure}
In Figure 4, we demonstrate the generalization of INDUS by calculating $P_v(N)$ in an arbitrarily shaped volume that is a collection of
non-overlapping sub-volumes. The volume that we have chosen spells, `I N D U S', using a collection of 156 cubic sub-volumes, each with a side of~$0.25\,$nm.

\begin{figure}[htbp] 
   \centering
   \vspace{-0.3in}
   \includegraphics[width=3.0in]{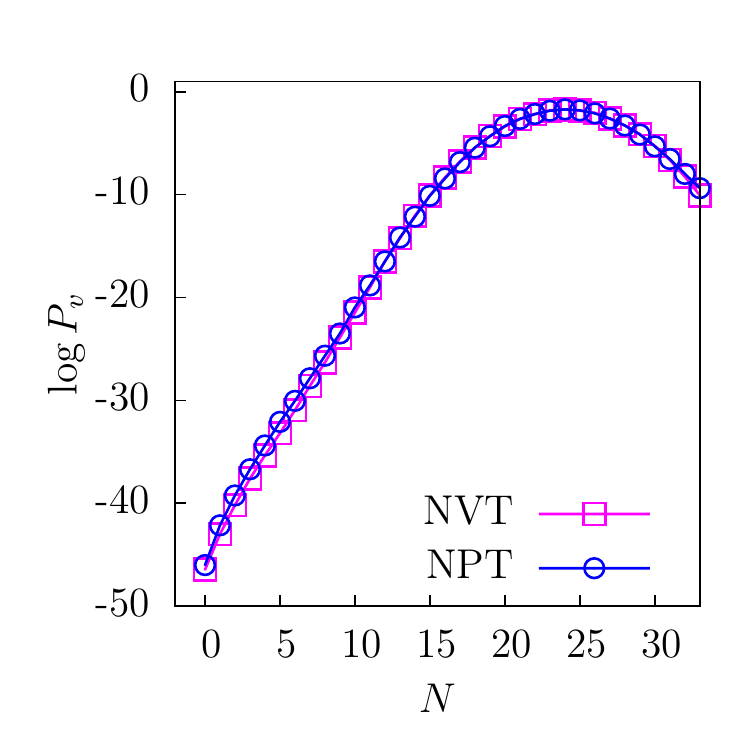}
   \caption{Comparing $P_v(N)$ for a cube of side~$0.9\,$nm, obtained using simulations in the NPT ensemble ($\mathcal{P}=1$bar) with that obtained from simulations in the NVT ensemble with a buffering vapor-liquid interface located far from~$v$.}
\end{figure}

In Figure~5, we show that for a cube of side~$0.9\,$nm, the $P_v(N)$
distribution calculated in the NPT ensemble at a pressure,~$\mathcal{P}=1\,$bar, is identical to that obtained in the NVT ensemble with a buffering vapor-liquid interface. This is expected since~$\mathcal{P}v \ll \kB T \ll \gamma A$, so the energetics of emptying~$v$ is governed almost entirely by the cost of forming an interface (Figure~3b). The effective pressure in the NVT system is the coexistence pressure, $\mathcal{P}^{*}$, at $T=300\,$K, which is close to~$0.06\,$bar. Since, $\mathcal{P}^{*}v < \mathcal{P}v \ll \kB T$, simulations in the NVT ensemble are an excellent approximation to NPT simulations at~$1\,$bar.

\begin{figure}[htbp] 
   \centering
   \includegraphics[width=3.0in]{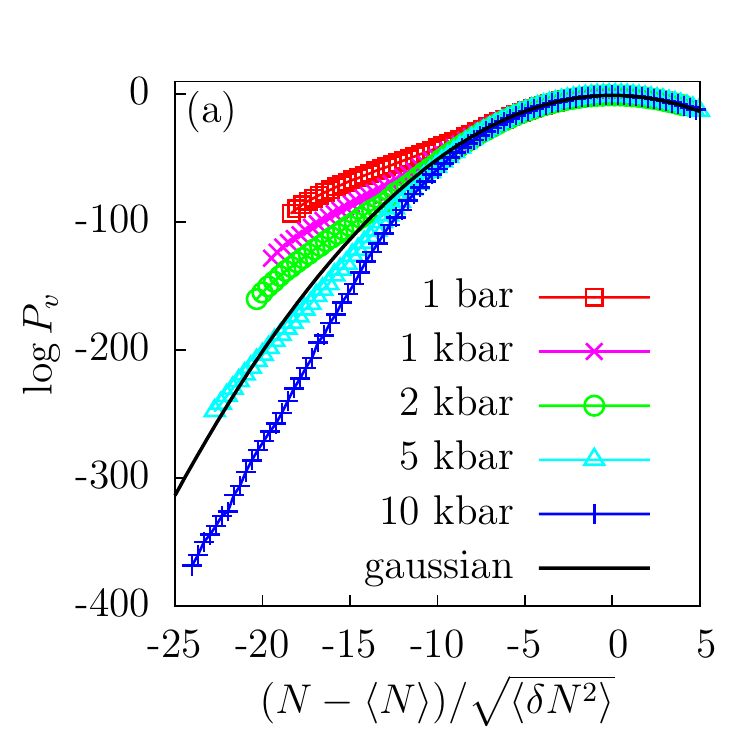}\includegraphics[width=3.0in]{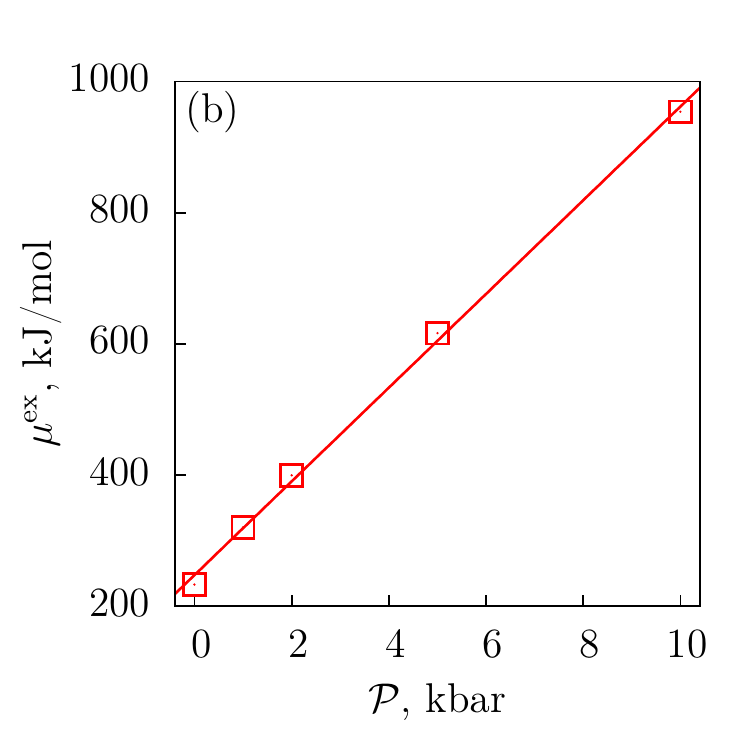}
   \caption{(a) $\log P_v$ as a function of $(N-\langle
   N\rangle)/\sqrt{\langle \delta N^2\rangle}$ for a cube of side~$1.2\,$nm, calculated in the NPT ensemble, over a range of
   pressures at $T=300\,$K.  (b) Free energy, $\mu^{\rm ex}$, of the same cube as a function of pressure. A
   linear fit yields the excess volume of the cavity, ${v^{\rm
   ex}}\approx0.67v$.}
\end{figure}

The ability to calculate $P_v(N)$ in the NPT ensemble allows us to
study its pressure dependence systematically. In Fig. 6a, we show
$P_v(N)$ distributions in a cube of side~$1.2\,$nm over a
broad range of pressures.  For pressures of~$1\,$kbar and higher, the~$\mathcal{P}v$ term is no longer negligible, and opposes
emptying $v$. Correspondingly, the low-$N$ fat tail disappears
gradually with increasing pressure. We also show in Fig.~6b that the free energy of hydrating the cubic cavity increases roughly linearly with pressure. The slope of $\mu^{\rm ex}$ versus $\mathcal{P}$ is the
excess volume for solvating the cavity, and is equal to $0.67v$ for this cubic probe volume.

\section{Conclusions}
Given the importance of density fluctuations in understanding a range
of solvation phenomena
\cite{garde96,hummer98pnas,netz2011,berne05_melittin,chu,berkowitz},
we anticipate that the INDUS method will be of broad interest. For instance, the size of density fluctuations at interfaces has been proposed recently as a
measure of interface hydrophobicity
\cite{AJP_Lscale,garde09pnas,acharya:faraday,garde09prl}. The extended
INDUS method is capable of characterizing hydrophobicity in
complex environments that exhibit chemical
heterogeneity~\cite{acharya:faraday,pgd06}, complex topography~\cite{wallqvist1,hummer_topo,luzar_fd}, and confinement~\cite{pgd06,berne_rev09,berne03,berne04,chou_dewet,pgd09}. The ability to calculate~$P_v(N)$ over a range of pressures using
the NPT ensemble will be useful in studying the effect of pressure on
biomolecular structure, and especially in quantifying the hydration
contribution to the pressure denaturation of proteins~\cite{ss_proteins}. Finally, quantifying $P_v(N)$ in a region
surrounding a solute molecule constitutes an important contribution in
the quasichemical theories of solvation~\cite{pratt_quasi,pratt_book}. Our extension of INDUS can be readily applied to quantify that contribution for a solute of arbitrary shape and size.

\begin{acknowledgements}
AJP would like to thank Sumanth Jamadagni for useful discussions. 
NIH Grant No. R01-GM078102-04 supported AJP in the early stages of this work, PV throughout, and DC in the early stages. In the later stages, DC was supported by the Director, Office of Science, Office of Basic Energy Sciences, Materials Sciences and Engineering Division and Chemical Sciences, Geosciences, and Biosciences Division of the U.S. Department of Energy under Contract No. DEAC02-05CH11231. SG gratefully acknowledges financial support of the NSF-NSEC (DMR-0642573) and NSF-CBET (0967937) grants.
\end{acknowledgements}


\end{document}